\documentclass[preprint,showpacs,preprintnumbers,amsmath,amssymb]{revtex4-1}
\usepackage{amssymb}
\usepackage{amsmath}
\usepackage{graphicx}

\begin{document}
\title{Coupled Spin-Phonon Excitations in Helical Multiferroics}

\author{ Chenglong Jia and Jamal Berakdar}
\affiliation{Institut f\"ur Physik, Martin-Luther-Universit\"at Halle-Wittenberg, 06099 Halle, Germany}

\begin{abstract}
Both the Dzyaloshiskii-Moriya interaction and the exchange-striction are shown to affect dynamically the magnetoelectric excitations in the perovskite multiferroic RMnO$_3$. The exchange-striction results in a biquadratic interaction between the spins and the transverse phonons, giving rise to quantum fluctuations of the ferroelectric polarization $\mathbf{P}$. This leads to  low-lying phonon modes that are perpendicular to  $\mathbf{P}$ and to the helical spins at small wave vector  but are  parallel to $\mathbf{P}$ at a wave vector close to the magnetic modulation vector. For spin-1/2 helimagnet, the local polarization can be completely reversed by the spin fluctuation, and so does the direction of the on-site spin chirality, which allows for a finite differential scattering intensity of polarized neutrons from a cycloidal magnet.
\end{abstract}


\maketitle
\section{Introduction}
Details of the coupling mechanisms of the magnetic and the ferroelectric order in multiferroics is currently under active research. This is due to the fundamental physics involved and  to promising  technological applications \cite{Multiferroics}.
Our focus here is the perovskite multiferroics  RMnO$_3$ with R = Tb, Dy, Gd and Eu$_{1-x}$Y$_x$ that have
 incommensurate spiral spin structure \cite{RMnO3}.  The experimental finding is that
RMnO$_3$ has a helical magnetic order and a finite ferroelectric (FE) polarization. The driving mechanisms of this ordering
is an interplay between the exchange interaction  and the Dzyaloshiskii-Moriya (DM) interaction.
Specifically, the spin-orbit coupling with a strength $\alpha$
 related to the $d(p)$-orbitals of the magnetic(oxygen) ions results
 in the FE  polarization  \cite{KNB,Jonh} $\mathbf{P} = \alpha \hat{e}_{ij} \times (\mathbf{S}_i \times \mathbf{S}_j)$.
   $\hat{e}_{ij}$ is a unit vector  connecting  the sites $i$ and $j$.
   Generally, it is to be  expected that the magnetoelectric (ME) coupling will affect not only the
    material static properties but also the dynamical response. Based on the spin-current model, the dynamical properties of DM interaction were studied in Ref.\cite{d-DM,d-GL,GdMnO3}. A novel magnon-phonon excitations so-called electromagnon, was theoretically predicted. When the spiral plane rotates with respected to the axis of the helical wave vector, so does the the induced electric polarization, which couples the magnetic excitation to the electric field $E$ of the radiation in the direction  perpendicular to the spin spiral plane \cite{d-DM}. Experimental observations in RMnO$_3$ \cite{ME-RMnO3} and Eu$_{0.75}$Y$_{0.25}$MnO$_3$ \cite{EuYMnO3-1} seems to be  consistent with this finding. However, a detailed  study of  the terahertz spectrum of Eu$_{1-x}$Y$_x$MnO$_3$ \cite{EuYMnO3-2} revealed that  infrared-absorption along the spontaneous polarization direction is also possible, which is not explained by theory mentioned above. This violation suggests that the static and the dynamic ME coupling may be different \cite{violation}. We carried out  a detailed investigations of  the dynamical properties of the multiferroics and  find  that both, the DM interaction and the (super)exchange striction play an essential role and need to be taken into account.
\section{Theoretical Model}
We consider a one-dimensional spin chain along the $z$-axis with a frustrated spin interaction. An effective model that captures the spin-phonon coupling \cite{d-DM,EPL} has the Hamiltonian

\begin{eqnarray}
&& H = H_s + H_{DM} + H_p \\
&& H_s = \sum_{\langle ij \rangle_{nn}} J_1(r_i-r_j) S_i \cdot S_j \nonumber\\
 && \quad \quad + \sum_{\langle lm \rangle_{nnn} } J_2(r_l-r_m) S_l \cdot S_m \nonumber \\
&& H_{DM} = - \lambda \sum_i \mathbf{u}_i \cdot [\hat{e}_z \times (S_i \times S_{i+1})] \nonumber \\
&& H_p = \frac{k}{2} \sum_i \mathbf{u}_i^2 + \frac{1}{2M} \sum_i \mathbf{P}_i^2 \nonumber
\end{eqnarray}
where the notation $\langle ij \rangle_{nn}$ indicates  nearest-neighboring (nn)  $i$ and $j$, and $\langle lm \rangle_{nnn} $ corresponds the next-nearest-neighboring (nnn) $l$ and $m$. The competition between the nn ferromagnetic interaction ($J_1<0$) and the nnn antiferromagnetic interaction ($J_2>0$) leads to magnetic frustration and  realizes a spiral spin ordering with the wave vector $\cos Q = -J_1/4J_2$ \cite{GdFeO3,LiCu2O2,NaCu2O2}. $H_p$ describe  optical phonons. The spin-phonon interaction $H_{DM}$ originates from a spin-orbital coupling and breaks the inversion symmetry along the chain.
Minimizing the energy yields the condition of the
atomic displacement and the local spin-configuration,
$\mathbf{u}_{i} =\frac{\lambda}{k}\hat{e}_{z}\times(
\mathbf{S}_{i}\times \mathbf{S}_{i+1})$.
%
%
Particularly, if the $zx$ helical spins along the chain, \emph{i.e.} $\mathbf{S}_i = S (\sin iQ , 0, \cos iQ)$, an uniform electric polarization $\mathbf{P}$ along the $x$ direction is induced by the condensation of the transverse optical (TO) phonons, $\mathbf{P}= e^{*} \mathbf{u}_0= - e^{*} \frac{\lambda S^2}{k} \sin Q \hat{e}_x$ with a Born charge $e^{*}$. Generally,  $\mathbf{u}_x$ cannot be softened through the hybridization between the TO phonons and the  magnons because of $k/M \gg JS$. The spontaneous FE polarization $\mathbf{P}_x$ is frozen  at $-e\mathbf{u}_0$ in the ferroelectric phase. However, after accounting for the superexchange striction, we have transverse acoustic (TA) phonons, which induces the fluctuation of the polarization hybridized with the spin bosons and soften thus the transverse phonon behavior.

Considering small atomic displacements perpendicular to the chain, $\mathbf{u}_i^{\perp} \cdot \hat{e}_z=0$, the exchange energy $J$ falls off as a power law  with the separation of the magnetic ions
\begin{eqnarray}
J_{1,2}(|\mathbf{r}_i-\mathbf{r}_j|)
&\approx&
J_{1,2}[1-{\gamma_{1,2} \over 2}(\mathbf{u}_i^{\perp} -\mathbf{u}_j^{\perp})^2]
\label{ES}
\end{eqnarray}
where $\gamma$  is in the range of $6-14$ \cite{Book}. The emerging TA phonon mode is coupled to the spins with the bi-quadratic interaction
%
$- \gamma_{1,2} J_{1,2} (\mathbf{u}_i^{\perp} -\mathbf{u}_j^{\perp})^2 (\mathbf{S}_i \cdot \mathbf{S}_j)$.
%
This dynamical coupling does not contribute any additional static electric polarization but induces the fluctuation of the electric dipole moment due to the low frequency excitation modes of TA phonon. We write explicitly the atomic displacements into two parts: (i) the statical part $\mathbf{u}_i = (u_0^x, 0, 0)$ and (ii) the dynamical part $\delta \mathbf{u}_i = (- \delta u_i^x, \delta u_i^y, 0)$. Retaining terms up to the second order in the quantum fluctuation, the spin-current model delivers the following coupling terms
%
\begin{eqnarray}
\tilde{H}_{DM} &=& -\lambda S \cos Q \sum_i \delta u_i^x (\tilde{S}_{i+1}^x - \tilde{S}_i^x) \nonumber \\ &~~~& - \lambda S \sum_i \delta u_i^y (\tilde{S}_i^y \cos Q_{i+1} - \tilde{S}_{i+1}^y \cos Q_i)
\label{dME}
\end{eqnarray}
in the rotated spin frame: $ S_{i}^{x}=\tilde {S}_{i}^{x}\cos iQ + \tilde{S}_{i}^{z}\sin iQ$, $S_{i}^{y}=\tilde{S}_{i}^{y}$, and $S_{i}^{z}=-\tilde{S}_{i}^{x}\sin iQ +\tilde{S}_{i}^{z}\cos iQ$.%

\section{Results and analysis}

In spin-1/2 multiferroics, such as LiCu$_2$O$_2$ \cite{LiCu2O2},  the spin fluctuations may spontaneously reverse  the local spin. Defining the vector of spin chirality as the average of the outer product of two adjacent spins $\hat{c}_i = (s_i \times s_{i+1})/|s_i \times s_{i+1}|$, in the RMnO$_3$-type multiferroics the direction of local FE polarization is determined by the on-site spin chirality.   The dynamical DM interaction Eq.(\ref{dME}) yields the coupling term  between the spin and the spin-chirality in the spin-1/2 multiferroics,
%
$\sum_i \hat{c}_i^x(\hat{s}_{i+1}^x - \hat{s}_i^x) = \sum_i \hat{s}_i^x (\hat{c}_{i-1}^x - \hat{c}_i^x)$,
%
which indicates that  when the spin at site $i$ is flipped, $\hat{s}_i \rightarrow -\hat{s}_i$, the direction of spin-chirality $\hat{c}_i$ and $\hat{c}_{i-1}$  are also reversed. Assuming all spins point along their corresponding classical directions in the ground state of the spin-1/2 helical magnet as in NaCu$_2$O$_2$, where a $J_1-J_2$ spin model provides a good description of the helix state \cite{NaCu2O2}. So the spin interaction can be ferromagnetically given as $-J_s(Q) \hat{s}_i \cdot \hat{s}_j$ where $Q$ is taken as the pitch angle along the chain. An effective model that describes the interplay between the helical spin and spin-chirality has the form
\begin{eqnarray}
H_{sc} = -\sum_{i,j} (J_s \hat{s}_i \cdot \hat{s}_j + J_{c} \hat{c}_i \cdot \hat{c}_j) - \gamma \sum_i \hat{s}_i^x (\hat{c}_{i-1}^x - \hat{c}_i^x).
\end{eqnarray}
  The Hilbert space can be considered as the tensor product space 
$
|i \rangle \rightarrow |s_i^z \rangle _s \otimes | c_i^z \rangle _c.
$
Now if the spin at site $i$ is flipped, the spin and spin-chirality excitations are mixed due to the spin-phonon coupling. The expected value of spin-chirality is given by
\begin{equation}
\langle \hat{c} \rangle =1 - \langle \hat{s} \rangle,
\label{E-value}
\end{equation}
which is less than one. The experimental data for a finite differential scattering intensity of polarized neutrons from LiCu$_2$O$_2$ \cite{LiCu2O2} suggests $\langle \hat{c} \rangle \approx 0.3$ which is consistent with the estimated value $\langle \hat{c} \rangle = 0.44$ based on the ordered moment, $0.56 \mu_B$ per magnetic copper site \cite{NaCu2O2}.

For magnet RMnO$_3$, the helical spin ordering occurs, corresponding to the condensation of the spin bosons. By using the standard linear-spin-wave approximation, a dynamical magon-phonon interaction reads,
\begin{eqnarray}
\tilde{H}_{DM} &=& -\lambda S \cos Q \sum_q \delta u_q^x \tilde{S}_q^x ( \cos q -1) \nonumber \\ &~~~& - \lambda S \sum_q \delta u_q^y \tilde{S}_{q\pm Q}^y ( e^{\mp i Q} - e^{i(q \pm Q)}) /2
\end{eqnarray}
$\delta u_q^y$ is hybridized with the spin at $q \pm Q$ (optical magnons), but $\delta u_q^x$ is coupled to $\tilde{S}^x$ at $q$ (acoustical magnons).
The  polarization correlation functions are given as
\begin{eqnarray*}
&& \ll \delta u_q^x | \delta u_{\bar{q}}^x \gg = \frac{\omega^2 - \omega_s^2}{M[\omega^4 - \omega^2(\omega_p^2 + \omega_s^2) + \omega_p^2 (\omega_{s}^2-\omega_{sp}^2)]}, \\
&& \ll \delta u_q^y | \delta u_{\bar{q}}^y \gg = \frac{1}{M[\omega^2 - \omega _p^2 +  \frac{\lambda^2 S^3}{2M}\sum_{q'=q \pm Q}G_{s}(q')]}.
\end{eqnarray*}
where $\omega_p$ is the frequency for the transverse phonon, $\omega_s (q)$ is the energy dispersion of the spin-excitation, $\omega'_{sp}(q)= [2(A(q)-2B(q))(\lambda ^2 S^3 \cos^2 Q (1- \cos q))/k']^{1/2}$, and $G_{s}(q \pm Q) = (A(q \pm Q) + 2 B(q \pm Q))(1- \cos (q \pm 2Q)/(\omega^2 - \omega_s ( q \pm Q))$ with
\begin{eqnarray}
A(q) &=& -J_1[\cos Q + \frac{1}{2}(1+ \cos Q)\cos q] \nonumber \\ &-& J_2[\cos 2Q + \frac{1}{2}(1+ \cos 2Q)\cos 2q] \nonumber \\ &+& \frac{\lambda^2 S^2 \sin^2 Q }{2k}(2-\cos q) \\
B(q) &=& {J_1 \over 4} (1-\cos Q) \cos q + {J_2 \over 4}(1- \cos 2Q) \cos 2q \nonumber \\ &-& \frac{\lambda^2 S^2  \sin^2 Q }{4k}\cos q
\end{eqnarray}
 At small wave vectors, $q \sim 0$ and $\omega_p \approx \sqrt{k/M}$, the TA phonon is decoupled from spins.
The antisymmetric DM interaction dominates over the spin-phonon coupling. $\delta u_0^y $ is coupled  via $(\tilde{S}_Q^y - \tilde{S}^y_{\bar{Q}})$  to the rotation of the spin plane and the direction of the polarization along the chain. However, at a wave vector close to the magnetic modulation vector, i.e. $q \sim Q$, both the symmetric and antisymmetric magnetoelectric interaction respond to the fluctuations of the polarization.  Especially, in the direction parallel to the FE polarization  $\mathbf{P}$, there is a low frequency range around  $\omega_{-}^x \cong \omega_s (Q)$ where $u^x$ couples  resonantly to light. Introducing an easy-plane spin anisotropy $D (S_y)^2$ into the spin system, we observe  nearly the same low-frequency behavior of the polarization correlation functions $\omega_{-}^x \approx \sqrt{JSD} \approx \omega_{-}^y$. These conclusions are also  qualitatively  consistent with experiment observations for Eu$_{1-x}$Y$_x$MnO$_3$ \cite{EuYMnO3-2}.

\section{Summary}
In conclusion, we studied the origin of the magnetoelectric dynamics in the orthorhombic perovskite RMnO$_3$. At a small wave vector, the DM interaction determines  the low-frequency behavior of the phonons. For a wave vector close to that of the magnetically modulated structure, the exchange striction induces fluctuations in the FE polarization, and additional low-lying mode parallel to the FE polarization  emerges.  Due to the  dynamical Dzyaloshiskii-Moriya interaction, the spin-chirality is strongly coupled to the spin fluctuation which implies a large quantum flucatuation of the spin-chirality  in the ordered spin-1/2 system and results in a finite   scattering intensity of polarized neutrons from a cycloidal helimagnet.

{\it Acknowledgement:}
This work is supported by the German Science Foundation DFG through  SFB762 -B7- {\it functionality of oxide interfaces}.

%
%

\end{document}